          [2001/04/25 1.1 (PWD)]
\documentclass[a4paper,twocolumn]{esapub2005} 
\usepackage{times}
\usepackage{natbib}
\usepackage[]{graphicx}

\def\PIFon{PIF$_{\rm on}$}
\def\PIFoff{PIF$_{\rm off}$}

\title{A tool for phase resolved spectroscopy with ISGRI}
\author{A. Segreto}
\author{C. Ferrigno}
\affil{Istituto di Astrofisica Spaziale (IASF-INAF), Palermo, Italy. E-mail: segreto@ifc.inaf.it}

\begin{document}

\keywords{INTEGRAL; ISGRI}

\maketitle

\begin{abstract}

INTEGRAL observations provide a large amount of data on accreting binary systems.

To study the spectral emission of these sources it is necessary to perform timing analysis and phase resolved spectroscopy, but these tasks are really cumbersome if performed with the tools based on the imaging extraction methods usually used for coded mask instruments.

Here, we present the software we have developed for the ISGRI instrument which allows to extract in a fast way light curves, pulse profiles, and phase resolved spectra, making data reduction a much easier task.

\end{abstract}

\section{Introduction}

Pulse-phase resolved analysis is a powerful method to understand the physical emission process in isolated or binary pulsars. To perform phase resolved spectroscopy with ISGRI \citep{Lebrun03} using standard tools it is however necessary to follow a cumbersome and considerable \mbox{time-consuming} procedure; we have then developed a dedicated software that allows to make the data reduction in a much faster and easier way. The software uses the pixel Photo Illumination Fraction (PIF) to extract scientific data.

Spectral extraction methods which use the PIF information have already been implemented for other coded mask instruments (e.g. in the official OSA software for JEM-X and in the standard extraction software for the BAT instrument on board SWIFT), but to apply this method to the ISGRI detector it is necessary to introduce a correction step to take into account the effects of its non-ideal imaging Point Spread Function (PSF).


\section{PIF Extraction Method}

Spectral extraction in coded mask instruments can be performed by analysing the sky images generated in different energy ranges.
Shadowgram deconvolution is however a time-consuming process and repeating it on many energy bins makes the whole spectral extraction process very slow. A much faster way to extract any scientific product (light curves, spectra, pulse profiles, etc.) when the source position is known, is by using the PIF information.

For each pixel, the PIF is the fraction of the pixel area illuminated by a source (i.e. PIF=1 if fully illuminated and PIF=0 if not illuminated at all).
Selected two illumination thresholds, \PIFon and \PIFoff, we can define two classes of detector pixels, based on their illumination by the source of interest:

\begin{eqnarray}
   \rm{ON} &:& \rm{PIF_{ij}} \geq \rm{PIF_{on}} \\
  \rm{OFF} &:& \rm{PIF_{ij}} \leq \rm{PIF_{off}}
\end{eqnarray}

For these two pixel classes, scientific products (light curves, spectra, pulse profiles, etc.) can be easily accumulated, like for a conventional collimated instrument; the background subtraction is obtained by the difference between the ON and OFF products, after proper normalization for the number of pixels in each class and, if necessary, for their efficiency.

The two illumination thresholds, \PIFon and \PIFoff , can be optimized so that the final product will have the best S/N ratio. It can be easily verified that for ISGRI, the optimum S/N ratio for background dominated sources is obtained when \PIFon $\approx 0.75$ and \PIFoff$ \approx 0.25$.

Since in this extraction method the shadowgram deconvolution and image analysis steps are avoided, the whole procedure is very fast and the processing times are limited only by disk I/O operations.

The outlined extraction method is however strictly valid when one single source is present in the instrument Field Of View (FOV) or when the imaging Point Spread Function (PSF) is ideal. In order to apply this method to \mbox{non-ideal} imaging detectors a correction step must be performed.

\section{Cross-Contamination}

\begin{figure}[!]
\centering
\includegraphics[width=7.5cm]{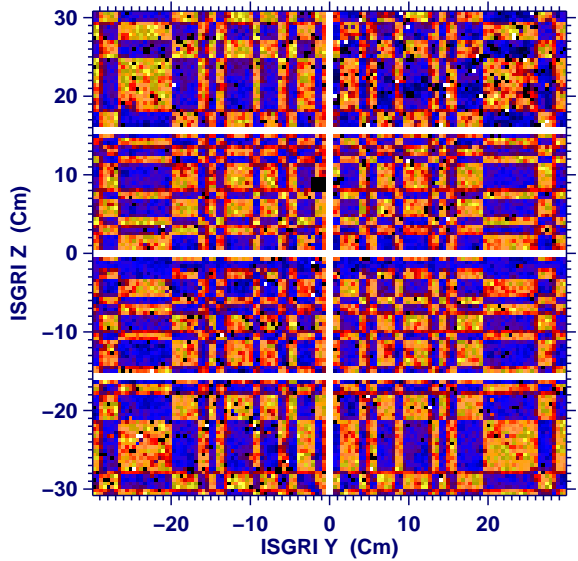}\\
\caption{ISGRI shadowgram for an almost on-axis source.
The gaps (two pixels wide) between the eight detector modules, are evident.
\label{fig:ISGRI_shadowgram}}
\end{figure}

The basic pattern of the IBIS coded mask \citep{Goldwurm03} has been designed to produce an imaging PSF  free of secondary lobes for sources observed in the Fully Coded Field Of View (FCFOV). The ISGRI detection plane however does not fully sample the mask shadow because of the gaps (two pixels wide) between the eight detector modules (Fig.~\ref{fig:ISGRI_shadowgram}).

\begin{figure}
\centering
\includegraphics[width=7.5cm]{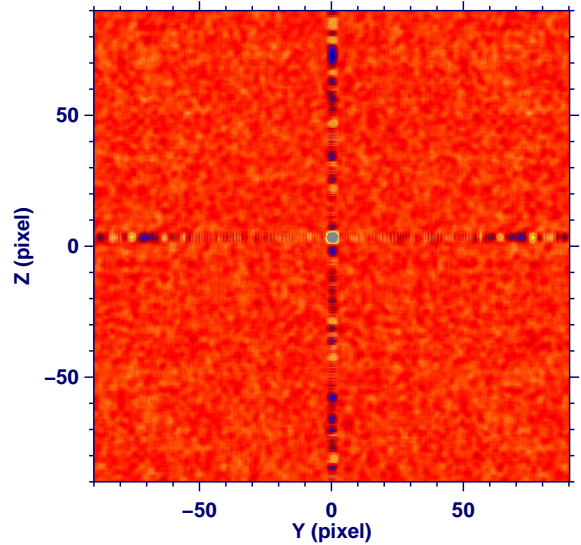}\\
\caption{ISGRI imaging PSF for an on-axis source. The numerous sidelobes forming a cross pattern are generated by the missing rows and column of pixel in the detection plane. The image show only the FCFOV so the eight main source ghosts are not visible.
\label{fig:ISGRI_PSF}}
\end{figure}

 Due to the missing rows and column of pixels in the \mbox{ISGRI} detector plane, even when a source is in the \mbox{FCFOV}, its PSF has numerous side lobes, forming a cross shaped pattern, that can reach intensity levels of several percentage of the main peak (Fig.~\ref{fig:ISGRI_PSF}). Dead pixels and a non-uniform distribution of the pixel efficiencies further contribute to the not ideal imaging characteristic of the instrument but, if randomly distributed, do not add other sidelobes to the PSF.

A non-ideal PSF implies that scientific products extracted by using the simple 'ON/OFF' background subtraction method may be affected by the presence of other sources in the same FOV. It is then necessary to compute and remove the effects of this 'cross-contamination'.

\subsection{Decontamination procedure}

If the positions of all sources in the instrument FOV is known, from their illumination patterns it is possible to compute the cross-contamination between any possible two of them. This information can be organized in a  square matrix C~($\rm N~\times~\rm N$) where N is the number of sources in the FOV. The diagonal elements, $c_{i,i}$, of this 'cross-contamination' matrix are the coding efficiency of the sources and the off-diagonal elements, $c_{i,j}$, the cross-contamination between two distinct sources.

The 'cross-contamination' matrix of an ideal coded mask instrument is of course diagonal since the illumination patterns of different sources are perfectly uncorrelated. For ISGRI  the off-diagonal elements assume non-zero values which depend on the relative source positions; in most cases they are of the order of 0.1\% but they can reach $10\%$ when a source is on one of the main sidelobes of the PSF of another source.
Higher values can be assumed when the PSF of a source partially overlaps with the PSF of another one (or with its ghost image).

In any case, the 'cross-contamination' matrix is not singular since on each row the highest values are on the diagonal ($c_{i,i} \approx 0.9$). The 'decontamination' matrix, \mbox {$D=C^{-1}$}, allows to compute the 'decontaminated' products as linear combination of the products generated by the ON/OFF background subtraction method.


\section{Software Verification}

\begin{figure}[!]
\centering
\includegraphics[width=5.9cm, angle=-90]{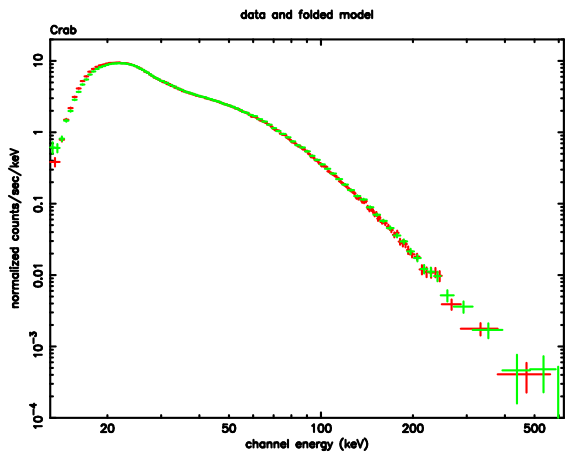}\\
\caption{
Comparison between the ISGRI Crab spectra extracted with the standard OSA pipeline and with the PIF based method. The two spectra are not distinguishable from each other.
\label{fig:Crab_spt_comparison}}
\end{figure}

\begin{figure}[!]
\centering
\includegraphics[width=5.9cm, angle=-90]{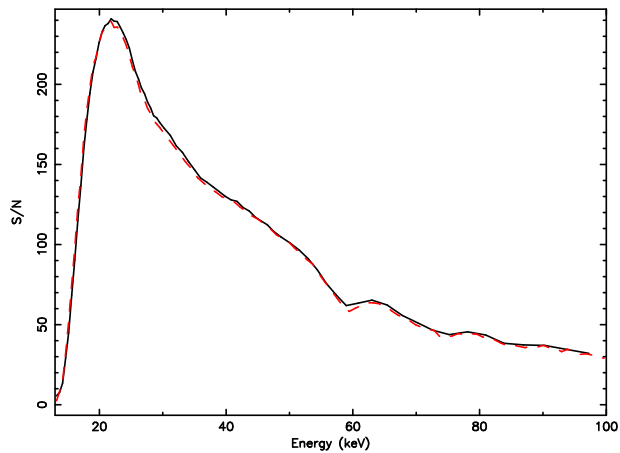}\\
\caption{
Comparison between the S/N of ISGRI Crab spectra extracted with the standard OSA pipeline and with the PIF based method illustrated in the text.
\label{fig:Crab_SN}}
\end{figure}

We have verified that the ISGRI spectra extracted with the standard OSA pipeline and with the software we have developed, based on the PIF method illustrated above, are equivalent in both shape and S/N ratio. In Fig.~\ref{fig:Crab_spt_comparison} and Fig.~\ref{fig:Crab_SN} the comparison between the Crab spectra and the relative S/N obtained with the two methods are shown.

\begin{figure}
\centering
\includegraphics[width=5.9cm,angle=-90]{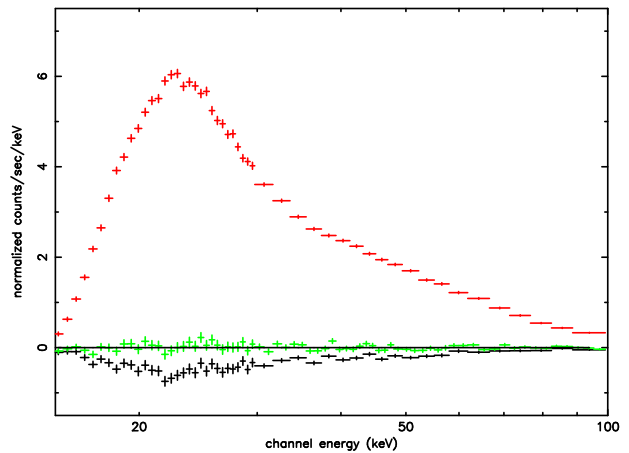}\\
\caption{Crab spectrum extracted with the PIF based method (the highest histogram) and spectra extracted in correspondence of the most negative sidelobe of its PSF before (the histogram with negative values) and after (the histogram consistent with zero) the application of the decontamination procedure.
\label{fig:negative_spectrum}}
\end{figure}

We have also verified the decontamination procedure by extracting the spectrum of a 'fake' source placed in correspondence of one of the main sidelobes of the PSF of a real source. As evident in Fig.~\ref{fig:negative_spectrum}, where a Crab observation has been used, the spectrum of the 'fake' source obtained without applying the decontamination procedure is not consistent with zero; this of course implies that the spectrum of a real source in that position would be severely altered.



After applying the decontamination procedure, the spectrum of the 'fake' source becomes consistent with zero, thus proving that the cross-contamination problem is solved and that the background has been correctly subtracted.


\section{Phase resolved spectroscopy}

When the source of interest is much stronger than all the other sources in the FOV, the cross-contamination can be neglected and phase resolved spectroscopy could be performed using standard software tools (developed for collimated telescopes) on the two events list files correspondent to the ON and OFF pixel classes.

\begin{figure}
\centering
\includegraphics[width=7.8cm]{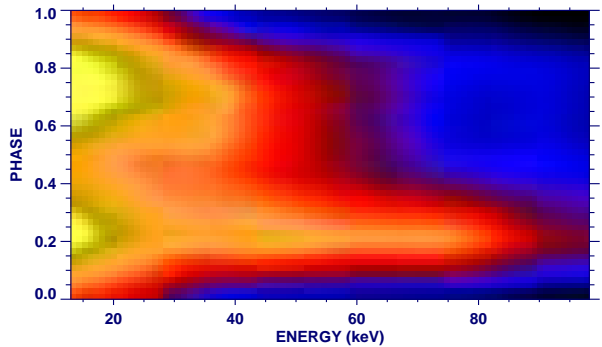}\\
\caption{Example of 'phase-energy' matrix for \mbox{1A 0535+262} observed with ISGRI. The evolution of pulse profile with energy is evident.
\label{fig:phase_energy}}
\end{figure}

In order to make the overall data reduction a much easier task than using standard tools, and to apply also the decontamination procedure, we have included in our software also the steps needed to perform phase resolved spectroscopy. From the spin period, its derivatives, and the eventual binary-orbital ephemeris of a pulsating source, the software computes for each event a 'phase tag' which is then used to accumulate a two-dimensional histogram: the 'phase-energy' matrix (like the one shown in Fig.~\ref{fig:phase_energy}).

Following the same procedure used to extract spectra and light curves, two 'phase-energy' matrices (for the ON and OFF pixel classes) are accumulated and then subtracted after proper normalization; if multiple sources are in the same field of view, the decontamination procedure is also applied.

Once the background subtracted 'phase-energy' matrices have been generated, phase resolved spectra and pulse profiles can be obtained by performing simple selections on its rows or columns. In this way, when it becomes necessary to change phase selections or energy binning, as it can often happen during the analysis of a pulsating source, it is not necessary to repeat the whole accumulation process again.

Since the accumulation process, even when performed with fine phase and energy binning, is very fast ($\approx 1$ sec. per scw.) and the extraction of any number of spectra or pulse profiles from the \mbox{'phase-energy'} matrix requires a negligible amount of time, the overall data extraction for phase resolved spectroscopy becomes easy and quick.


\section{Conclusions}

We have developed a software tool for the ISGRI instrument that allows to extract spectra, light curves and perform phase resolved spectroscopy of pulsating sources in a fast and easy way.

The extraction method is based on the PIF information, but it has been modified to take into account the possible cross-contamination between sources in the same FOV caused by the non-ideal imaging PSF of the detector. We have shown, in fact, that in some cases the cross-contamination plays an important role and that it can be computed and removed performing simple algebraic operations.



We have verified that the scientific products obtained  with our software are equivalent to the ones produced with the ISGRI standard analysis (OSA) in both shape and S/N ratio, but they are generated in much less time. In fact, since the scientific products are generated by performing simple algebraic operations directly on the event list, without imaging analysis, the accumulation time, for a typical INTEGRAL science window, is reduced to the order of few seconds (limited by the data I/O operations).

Moreover, since the processing time is independent on the number of energy or time bins used, spectra and light curves can be computed at the maximum resolution and binned successively, depending of the desired S/N ratio. In this way it is not necessary to repeat  the whole accumulation process again (as needed when using the standard analysis software) if one should decide to change the size of the energy or time bins.

Our software includes also tools dedicated to the phase resolved spectroscopy of pulsating sources. In particular it is possible to fold the data with any spin period and derivatives and, if necessary, correct the event arrival time to binary system barycenter. The results of the accumulation process are background subtracted \mbox{'phase-energy'} matrices from which phase resolved spectra and pulse profiles can be later extracted, with a simple selection software.

Example of pulsating sources observed with ISGRI and whose phase resolved analysis has been performed with our software \footnote{\texttt{the sowtware may be obtained on request from authors}} are 1A 0535+262 \citep{Cabalero06}, 1E1145.1-6141 \citep{Ferrigno06}, and GX 1+4\citep{Ferrigno206}.


\section*{Acknowledgments}

We wish to thank T. Mineo for many useful comments and suggestions.


\begin{thebibliography}{}

\bibitem[Ubertini(2003)]{Ubertini03}
Ubertini, et al. 2003, A\&A, 411, L131, 2003


\bibitem[Lebrun et al.(2003)]{Lebrun03} Lebrun, F., et al.\
2003, A\&A, 411, L141

\bibitem[Goldwurm(2003)]{Goldwurm03}
Goldwurm, A. et al., A\&A, 411, L223, 2003


\bibitem[Caballero(2006)]{Cabalero06}
Caballero I. et al., these proceedings

\bibitem[Ferrigno(2006)]{Ferrigno06}
Ferrigno C. et al., these proceedings


\bibitem[Ferrigno et al.(2007)]{Ferrigno206} Ferrigno, C., Segreto,
A., Santangelo, A., Wilms, J., Kreykenbohm, I., Denis, M., \& Staubert, R.\
2007, A\&A, 462, 995




\end{thebibliography}

\end{document}